\documentstyle[prl,aps,epsf,multicol]{revtex}
\begin{document}
\title{Conventional mechanisms for ``exotic'' superconductivity}
\author{D.F. Agterberg$^{\dagger}$, Victor Barzykin$^{\dagger}$, 
and Lev P. Gor'kov$^{\dagger,*}$}
\address{$^{\dagger}$National High Magnetic Field Laboratory, 
Florida State University,\\
1800 E. Paul Dirac Dr., Tallahassee, Florida 32310 \\
and \\
$^*$L.D. Landau Institute for Theoretical Physics,
Chernogolovka, 142432, Russia
}
\maketitle
\begin{abstract}
We consider the pairing state due to
the usual BCS mechanism in substances of cubic and hexagonal symmetry
where the Fermi surface
 forms pockets around several points of high symmetry.
We find that the symmetry imposed on the multiple pocket positions  
could give rise
to a multidimensional nontrivial 
superconducting order parameter. The time reversal symmetry in the pairing
state is broken. We suggest several candidate substances where such 
ordering may appear, and discuss means by which such a phase 
may be identified.
\end{abstract}
\begin{multicols}{2}
\narrowtext
Most conventional superconductors are described very well by the BCS 
theory\cite{BCS}.
The electron-phonon interaction mediates an attraction between electrons
that is stronger than the Coulomb repulsion. This  
gives rise to the Cooper instability of the
normal state leading to the appearance of a condensate of pairs. 
The order parameter
(the anomalous function $\cal{F}$ \cite{gorkov}) in this case belongs to the "s-wave"
type, {\it i.e.}, it is invariant with respect to the transformations of $G \otimes R$,
where $G$ is the crystal point group and $R$ is time reversal operation. 
As a result the
quasiparticle spectrum has a gap, 
which leads to well-known experimental consequences.
A variety of materials: $^3$He \cite{he3}, UBe$_{13}$ \cite{ott}, UPt$_3$
\cite{ste84}, high $T_c$
materials \cite{bed86}, and Sr$_2$RuO$_4$\cite{mae94}, 
have been discovered that potentially
break the $G \otimes R$ symmetry of the normal state. A well known 
example is $A$-phase of $^3$He \cite{he3} which is 
not rotationally or time reversal invariant (note that the $B$ phase
of $^3$He is both rotation and time reversal invariant). 
Such non $s$-wave superconductors are usually expected to 
have a gapless excitation 
spectrum and arise when the interaction itself
depends upon the superconducting ground state (in the case of $^3$He
the BCS ground state is the $B$ phase and the spin fluctuation feedback
effect is required to stabilize the $A$ phase \cite{Anderson}). 
All the possible symmetry classes
of the superconducting state in crystalline materials were enumerated 
in Ref.\cite{vg} (for a review see Ref.\cite{gor,ru}). 

We show below that exotic superconductivity can be a much more 
common phenomenon and does not require  
unusual mechanisms. The 
electron-phonon and Coulomb interactions are enough to give rise 
to an multidimensional  order
parameter which would have lower symmetry than the normal state -  
including the breaking
of time-reversal invariance. 
The effects we consider are possible in metals with several pockets
which are centered at or around some symmetry points of the Brillouin 
Zone (BZ). 
A BCS approximation generalized to the the multi-band case
(see, {\it e.g.}, \cite{suh59}) will be used. 
The new point here is that since the form of the interaction parameters  
describing the  two electron    
scattering on and between the different pockets of the Fermi surface 
(FS) is fixed
by symmetry, the resulting superconducting state 
need not be $s$-wave.    
Below we consider three cases in detail:
a) three FS pockets centered about the 
X-points of a simple cubic lattice; 
b) three FS pockets at the M-points of the hexagonal
lattice; c) four FS pockets at the L-points in the face-centered cubic 
lattice. A complete analysis of all other high symmetry points is possible,
and will be published elsewhere\cite{abg}.

\begin{figure}
\epsfxsize=3.5 in
\epsfbox{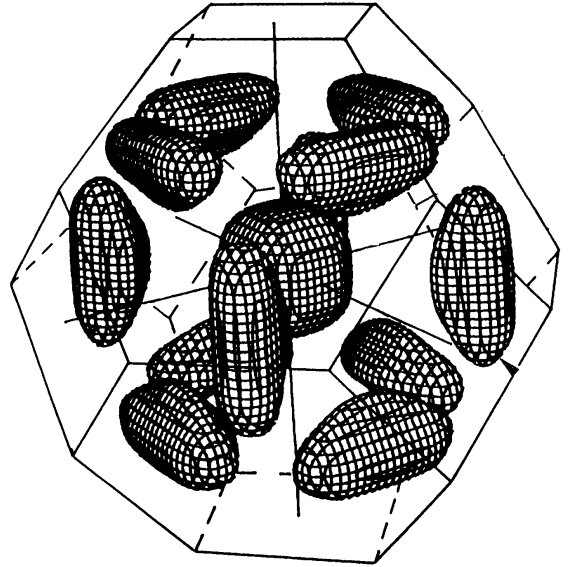}
\caption{Some FS sheets of CeCo$_2$
(from Ref.\protect{\cite{sug96}}).} 
\label{fig1}
\end{figure}

We emphasize that this FS structure is not unusual. Indeed, 
superconductivity with pockets as in case (a) and $T_c \sim 0.1K$ is
found in LaB$_6$\cite{LaB6}. Another example is given by the superconducting
semiconductors such as PbTe, SnTe, or SrTiO$_3$ \cite{cohen}. Many  
materials exist where 
such FS sheets  co-exist with other non-symmetry 
related FS sheets and some of these materials 
have anomalous superconducting
properties. One example is CeCo$_2$ \cite{sug95,sug96},  
 Fig.~\ref{fig1} shows some of the FS sheets
of CeCo$_2$ ($T_c=1.6$ K).
We will return
to this later\cite{abg}.

We will use the generalized Ginsburg-Landau (GL) functional  
to identify possible nontrivial superconducting phases. 
The Hamiltonian for several separate pieces of the FS can be
written in the following form:
\end{multicols} \widetext
\begin{equation}
H= \sum_{\alpha \sigma {\bf p}} \epsilon({\bf p}) 
a^{\dagger}_{\alpha \sigma}({\bf p}) a_{\alpha \sigma}({\bf p}) 
+ {1 \over 2} \sum_{{\bf k},{\bf k'},{\bf q}} 
\sum_{\alpha \beta \sigma \sigma'} \lambda_{\alpha \beta}({\bf q}) 
a^{\dagger}_{\alpha \sigma}({\bf k+q}) a^{\dagger}_{\beta \sigma'}({\bf k'-q})
a_{\alpha \sigma'}({\bf k'}) a_{\beta \sigma}({\bf k}),
\end{equation}
\begin{multicols}{2} \narrowtext
\noindent where  $\sigma$ and $\sigma'$ are
spin indices, $\lambda_{\alpha \beta}({\bf q})$ includes the interaction
for scattering two electrons from the pocket $\alpha$ into pocket $\beta$ 
which is due to both Coulomb and electron-phonon terms. 
Introducing the anomalous Green's function 
$\hat{\cal{F}}_{\alpha}(x-x')$ for each FS sheet $\alpha$, 
the corresponding Gor'kov equations\cite{AGD} can be used to 
obtain the following
solution at finite temperatures for the case of singlet pairing:
\begin{equation}
{\cal{F}}_{\alpha}^{\dagger}(\omega_n,{\bf p}) = {\Delta_{\alpha}^*({\bf p}) \over \omega_n^2 +
\xi^2 + |\Delta_{\alpha}({\bf p})|^2},
\label{gap0}
\end{equation}
where
\begin{equation}
\Delta^*_{\alpha}({\bf p}) = - {T \over (2 \pi)^3} \sum_{\beta}\sum_n \int
d{\bf k} \lambda_{\beta \alpha}({\bf p - k}) F^{\dagger}_{\beta}(\omega_n, {\bf k}).
\label{gap1}
\end{equation}
Eq.(\ref{gap0}) being expanded in $|\Delta_{\alpha}|$ up to the third
order, Eq.(\ref{gap1}) becomes a variation of the GL
functional with respect
to the vector order parameter $\Delta_{\alpha}({\bf p})$\cite{GL,GLG}. 
For simplicity we assume 
that each $\Delta_{\alpha}({\bf p})$ 
is constant along the corresponding FS, and  
to fourth order in $\Delta_{\alpha}$ we can write
\end{multicols} \widetext
\begin{equation}
F_s - F_n = - \Delta^*_{\alpha} \left[(\hat{\lambda}^{-1})_{\alpha \beta} - 
\delta_{\alpha \beta}{ m p_0 \over 2 \pi^2} 
\ln\left({2 \gamma \omega_D \over \pi T}\right)\right] \Delta_{\beta} 
+ {7 \zeta(3) m p_0 \over 32 \pi^4 T^2} \sum_{\beta} |\Delta_{\beta}|^4
\label{F4}
\end{equation} 
\begin{multicols}{2} \narrowtext
\noindent where $\hat{\lambda}^{-1}$ is the matrix inverse to the interaction 
$\lambda_{\alpha \beta}$, $\omega_D$ is the cutoff (Debye) frequency.

We now analyze three different cases for multiple 
FS sheets. 

\bigskip
(a)  {\em Three X points in a cubic lattice}.

The interaction matrix $\hat{\lambda}$ for three X points takes the 
following general form:
\begin{equation}
\lambda_{\alpha \beta} = \lambda \delta_{\alpha \beta} + \mu (1-\delta_{\alpha \beta}).
\label{int}
\end{equation}
Here $\lambda$ is the interaction on the same pocket, $\mu$ couples any 
two different pockets.
Consider first the linearized gap equation Eqs.(\ref{gap0}) and (\ref{gap1}) 
to determine $T_c$:
\begin{equation}
\Delta_{\alpha}^* {2 \pi^2 \over m p_0} = - \sum_{\beta} \lambda_{\alpha \beta}
\Delta_{\beta}^* \ln\left({2 \gamma \omega_D \over \pi T_c}\right)
\label{Tcc} 
\end{equation} 
The three
$\Delta_{\alpha}$ transform among  each other at cubic symmetry transformations
forming a 3D reducible representation of the cubic 
group $O_h$, which {\em is split} into a $1D$ $A_{1g}$ and 
a $2D$ $E_g$ irreducible representation.
These two representations correspond to different order
parameters with two critical temperatures:
\begin{eqnarray}
T_{c,E} &=& {2 \gamma \omega_D \over \pi} 
\exp\left({2 \pi^2 \over m p_0 (\lambda - \mu)}\right) \hspace{1cm} (2D) \\
T_{c,A} &=& {2 \gamma \omega_D \over \pi}
\exp\left({2 \pi^2 \over m p_0 (\lambda + 2 \mu)}\right) \hspace{1cm} (1D)
\label{TC} 
\end{eqnarray}
(the terms in the exponents must be  negative for the 
Cooper effect to take place). 
The basis wave function for 1D identical representation is
\begin{equation}
l = (\Delta_1 + \Delta_2 + \Delta_3)/\sqrt{3}
\label{1Dr}
\end{equation}
and the basis wave functions for the 2D representation can be chosen as
\begin{eqnarray}
\eta_1 &=& (\Delta_1 + \epsilon \Delta_2 + \epsilon^2 \Delta_3)/\sqrt{3} \nonumber \\
\eta_2 &=& (\Delta_1 + \epsilon^2 \Delta_2 + \epsilon \Delta_3)/\sqrt{3},
\label{2Dr}
\end{eqnarray}
where $\epsilon = \exp(2 \pi i/3)$.
From Eq.(\ref{TC}) if $\lambda - \mu < 0$ and $\mu > 0$ then superconductivity  
will belong to the nontrivial 2D $E_g$ representation,
 {\it i.e.}, if the interaction {\em between two} different 
FS pockets is dominated by Coulomb repulsion.  
Let us consider the
latter case in detail. Rewriting the Landau functional Eq.(\ref{F4}) in terms of $l$, $\eta_1$ 
and $\eta_2$, we obtain, for temperatures $T$ {\it near} $T_{c,E}$:
\end{multicols} \widetext
\begin{equation}  
{2 \pi^2 \over m p_0} \delta F = {T-T_{c,E} \over T_{c,E}} (|\eta_1|^2+|\eta_2|^2) + \ln(T_{c,E}/T_{c,A})|l|^2 +
{7 \zeta(3) \over 48 \pi^2 T_{c,E}^2}(|\eta_1|^4 + |\eta_2|^4+ 4 |\eta_1|^2 |\eta_2|^2
+ F^{(4)}_{l \eta}),
\label{F2D} 
\end{equation}
\begin{multicols}{2} \narrowtext 
\noindent where $F^{(4)}_{l \eta}$ is  the fourth order term in the GL
functional
which may admix the $1D$ representation $l$:
\begin{eqnarray}
F^{(4)}_{l \eta} =
 2 l (\eta_1^*)^2 \eta_2+ 2 l \eta_1 (\eta_2^*)^2 + h.c.
\label{termsl}
\end{eqnarray}
With $T_{c,A} < T_{c,E}$  the superconducting instability will correspond to
the $2D$ representation $E_g$ of the cubic point group. The fourth  order
coefficients in Eq.(\ref{F2D}) indicate that the class $O(D_2)$ is the most preferable 
energetically\cite{vg}. 
This class corresponds to a phase with $\eta_2=0, \ \ \eta_1 \neq 0$ in Eq.(\ref{F2D}).
The symmetry properties of this class are known\cite{vg}.
Time-reversal symmetry is broken and allows for 
antiferromagnetic domains and for fractional vortices to appear 
(see, {\it e.g.}, \cite{ru}). 
In principle point nodes  
should appear where the FS intersects the cube diagonals. 
This would lead to the {\it electronic} contribution 
in the $T^3$ behavior for the heat capacity at low
temperatures. In our case, however, there is no 
FS along the diagonals of the 
cube 
and  the low temperature thermodynamic properties will be determined by 
the gap of the same magnitude for all three FS sheets. 
In the presence of another FS, for example, at the 
$\Gamma$-point,
the nontrivial order $O(D_2)$ will be induced on it.
In this case the point nodes will exist and power laws in thermodynamic
properties due to the superconductivity should  be seen experimentally.
The anisotropy of the upper critical field ($H_{c2}$) \cite{gor84} 
{\it near} $T_c$   
for this class also requires that (at least) two vortex lattice phases
(with a second order transition between them) exists when the magnetic 
field is applied along the $(1,1,0)$ and 
equivalent directions. 
Note from Eq.(\ref{termsl}) that
the terms linear in $l$ identically disappear for this class, 
{\it i.e.}, there will be no admixture of
the $s$-wave component.  

(b)  {\em Three X points in a hexagonal lattice}.

Calculations in this case are the same as in the previous case, {\it i.e.},
Eqs.(\ref{int})-(\ref{termsl}) apply. 
We only have to specify the symmetry properties
and the superconducting class in this case since 
the symmetry of the lattice is different.
The three-dimensional representation is split by the hexagonal group 
$D_{6h} \otimes R$ into 1D($A_{1g}$) and 2D($E_{2g}$). Note that the basis functions
for $E_{2g}$ can be once again chosen as given 
by Eq.(\ref{2Dr}). The phase with 
$\eta_1 \neq 0, \ \ \ \eta_2=0$ has the lowest 
free energy and in this case corresponds to the nontrivial class $D_6(C_2)$. Time-reversal symmetry
for this class is broken\cite{vg}, and {\em ferromagnetism} is allowed. 
Point nodes (at two points of intersection of an additional FS at the $\Gamma$-point
with the six-fold axis) can be seen in thermodynamic properties but again
these nodes are only present if such a FS exists.
The upper critical field is isotropic near $T_c$ for this superconductivity 
representation. Nevertheless, it can be shown that 
there will also exist (at least) 
two distinct vortex lattice phases
with a second order transition between them for the magnetic field applied
in the basal plane \cite{abg}.

\bigskip
\noindent
(c)  {\em Four L-points in the fcc lattice.}

The interaction and the linearized gap equation for the four $L$ points again 
take the form Eqs.(\ref{int}-\ref{Tcc}). This time the 4D representation 
$\Delta_{\alpha}$ is split into the 
1D($A_{1g}$) and 3D($F_{2g}$) irreducible representations.
\begin{figure}
\epsfxsize=3 in
\epsfbox{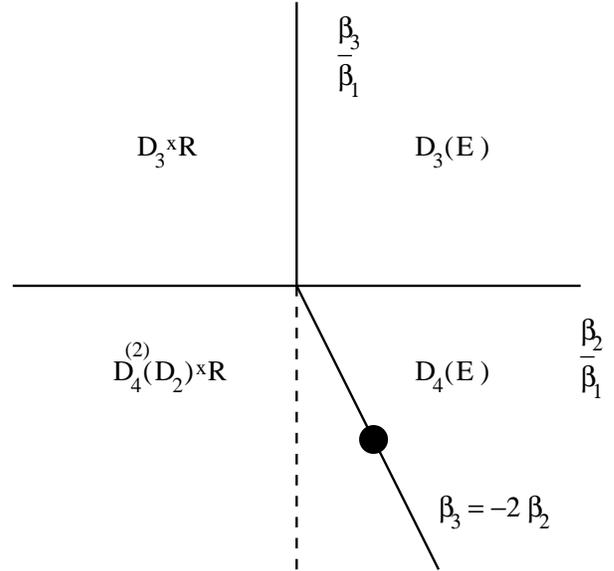} 
\caption{Regions of existence of different superconducting phases on the
basis of the three-dimensional representations of the cubic group
(from Ref.\protect{\cite{vg}}). A point on the boundary 
of two phases corresponds to
our BCS solution.} \label{phases}
\end{figure}
The critical temperatures for the two 
representations are now given by
\begin{eqnarray}
T_{c,F} &=& {2 \gamma \omega_D \over \pi} 
\exp\left({2 \pi^2 \over m p_0 (\lambda - \mu)}\right) \hspace{1cm} (3D) \\
T_{c,A} &=& {2 \gamma \omega_D \over \pi}
\exp\left({2 \pi^2 \over m p_0 (\lambda + 3 \mu)}\right) \hspace{1cm} (1D),
\label{TC1} 
\end{eqnarray}
The basis functions for the 1D and 3D representations are 
\begin{equation}
l = (\Delta_1 + \Delta_2 + \Delta_3 + \Delta_4)/2 \ \ \ \ \ (1D)
\label{1Dr'}
\end{equation}
and
\begin{eqnarray}
\eta_x &=& (\Delta_1 - \Delta_2 - \Delta_3 + \Delta_4)/2 \nonumber \\
\eta_y &=& (\Delta_1 + \Delta_2 - \Delta_3 - \Delta_4)/2 \ \ \ \ \ (3D) \label{3Dr} \\
\eta_z &=& (\Delta_1 - \Delta_2 + \Delta_3 - \Delta_4)/2 \nonumber.
\end{eqnarray}
The nontrivial 3D representation is stable if $\lambda - \mu < 0$ and $\mu > 0$,
i.e. if the interaction is {\em attractive} for each pocket alone, while
it is {\em repulsive} between two different pockets. 
As above, we can expand Eq.(\ref{F4}) in terms of $l$ and $\vec{\eta}$.
Dropping the 1D identical representation, we get:
\end{multicols} \widetext
\begin{equation}  
{2 \pi^2 \over m p_0} \delta F = {T-T_{c,F} \over T_{c,F}} (\vec{\eta}\vec{\eta}^*) + 
{7 \zeta(3) \over 64 \pi^2 T_{c,F}^2}[  2 (\vec{\eta}\vec{\eta}^*)^2 + 
|\vec{\eta}^2|^2 - 2 (|\eta_x|^4 + |\eta_y|^4+ |\eta_z|^4)].
\label{F3D} 
\end{equation}
\begin{multicols}{2} \narrowtext 
The GL coefficients in Eq.(\ref{F3D}) places the 
system right on the boundary of two phases, superconducting classes
$D_4^{(2)} (D_2) \otimes R$ and $D_4(E)$ (see Fig.\ref{phases}). This degeneracy
is an artifact of the BCS theory, it is {\em not} lifted by higher order terms
in the GL functional.
The presence of a FS at the $\Gamma$ point  
lifts this degeneracy. As a result, the magnetic 
superconducting class $D_4(E)$ is likely to appear. 
This class also allows ferromagnetism.
Note that it has  a line of nodes (at the intersection of the FS 
with the horizontal plane of symmetry) 
{\it i.e.}, $C \propto T^2$ at low temperatures\cite{vg}.
There are also point nodes at the
intersection of the FS with the four-fold symmetry axis. 
This representation also exhibits an anisotropy of $H_{c2}$ near $T_c$.
In principle multiple vortex lattice phases can also exist for this
class but they are not required by symmetry as they are for the $E$ 
representations discussed above.

In the above cases, other than the 
standard isotropic order parameter, only 
multidimensional order parameters appeared. This leads to the possibility
of domain walls between different equivalent superconducting states
and as a possible consequence the existence of inhomogeneous magnetic order
\cite{vg} (see also \cite{gor,ru}). 
Also in all the above cases the
resulting superconducting states had gaps of equal magnitude on each
of the FS sheets. In such a case 
a Hebel-Slichter peak in $1/T_1$ measurements may be present. The 
non-trivial representations were stable when the pair interaction between
the different sheets was repulsive (independent of the intra-sheet 
interaction).  
In many materials the Coulomb repulsion can be
comparable to the attraction between
electrons due to electron-phonon interactions.
This is illustrated through the reduction of isotope effect
due to Coulomb repulsion.
It is well known that
for a number of metals $T_c \propto M^{- \alpha}$,
where not only $\alpha \neq 0.5$ but 
it may even have the opposite sign.
($\alpha \simeq -5$ for $\alpha-U$ \cite{par}).
The sensitivity of these exotic superconducting phases to impurities
needs a more detailed analysis. However, provided
the inter-pocket defect scattering amplitudes are much smaller than the 
intra-pocket amplitudes, these phases will survive the presence 
of a considerable amount of defects (due to the ordinary BCS pairing
on each sheet). This will be studied in more detail in \cite{abg}.

In summary, we have shown that exotic superconductivity 
can appear merely as a competition of the phonon and Coulomb interactions
if the FS consists of several pockets located at some 
symmetry points. Time-reversal symmetry is broken for the nontrivial 
order, meaning that the superconducting transition should be 
accompanied by  some kind of magnetic order. 
The simplest methods to identify exotic order
parameters are, apart from the phase-sensitive measurements,
the power law dependence of the heat capacity (due to the nodes),  
measurements of the upper critical field anisotropy $H_{c2}(\theta)$ at $T_c$
\cite{gor84}, or the observation of transitions between different
vortex lattice phases. 
Note that if the FS pockets  
are fully isolated then the nodes are absent  
since the order parameter is then  constant
on each FS pocket and changes phase as one moves 
from one pocket to another.  Nodes could appear, however, if there are
``necks'' connecting different sheets \cite{LaB6} or if superconductivity 
is induced on a FS centered, for example, 
around the $\Gamma$-point. The upper critical field anisotropy
 near $T_c$ does not work as a test of nontrivial order
in the hexagonal group \cite{gor84}. The
magnetic order, on the other hand, can  be observed in 
$\mu$SR measurements or magnetization measurements in small enough samples
(where the dimensions are on the order of the penetration depth). A 
general classification
of all cases of nontrivial superconductivity of the type considered
above is possible. These FS sheets 
are not always centered
on the BZ boundary, as, for example, in some doped
semiconductors\cite{cohen} and CeCo$_2$ \cite{sug96}. 
We postpone a detailed analysis 
of the various possibilities to a future work\cite{abg}.

We would like to thank Z.Fisk, D. Khokhlov, J.R. Schrieffer, 
and the members
of the NHMFL condensed matter theory group seminar for
useful discussions and comments.
This work was supported by the National High Magnetic
Field Laboratory through NSF cooperative agreement 
No. DMR-9527035 and the
State of Florida.

\end{multicols} 
\end{document}